\documentclass[10pt,conference,onecolumn]{IEEEtran}
\usepackage{epsfig}
\usepackage{amsmath}
\usepackage{amstext}
\usepackage{amsfonts}
\usepackage{amssymb}
\usepackage{eucal}
\usepackage{graphicx}
\usepackage{verbatim}
\usepackage{stmaryrd}
\usepackage{pstricks}
\usepackage{pst-node}
\usepackage{psfrag}

\usepackage{tikz}
\usepackage{algorithmic}

\newcommand{\EE}{\mathrm{E}}

\DeclareMathOperator{\diag}{diag}

\begin{document}
\bibliographystyle{IEEEtran}

\title{A maximum entropy approach to OFDM channel estimation}

\author{\authorblockN{Romain Couillet}
\authorblockA{NXP Semiconductors, Supelec\\
505 Route des Lucioles\\
06560 Sophia Antipolis, France\\
Email: romain.couillet@nxp.com}
\and
\authorblockN{M{\'e}rouane~Debbah}
\authorblockA{Alcatel-Lucent Chair, Supelec\\
Plateau de Moulon, 3 rue Joliot-Curie \\
91192 Gif sur Yvette, France\\
Email: merouane.debbah@supelec.fr}
}

\maketitle

\begin{abstract}
In this work, a new Bayesian framework for OFDM channel estimation is proposed. Using Jaynes' maximum entropy principle to derive prior information, we successively tackle the situations when only the channel delay spread is a priori known, then when it is not known. Exploitation of the time-frequency dimensions are also considered in this framework, to derive the optimal channel estimation associated to some performance measure under any state of knowledge. Simulations corroborate the optimality claim and always prove as good or better in performance than classical estimators.
\end{abstract}

\section{Introduction}
Modern high rate wireless communication systems, such as IEEE-Wimax \cite{WiMax} or 3GPP-Long Term Evolution (LTE) \cite{LTE}, usually come along with large bandwidths. In multipath fading channels, this entails high frequency selectivity, which theoretically is beneficial for it provides system diversity. But in practice, this constitutes a strong challenge for equalization. The orthogonal frequency division multiplexing (OFDM) modulation \cite{BIN90}, considered as the scheme for most future wireless systems, allows for simplified equalization through pilot sequences scattered in the time-frequency grid and possibly over the space dimension when multiple antennas are used.

The challenge in channel estimation with a limited number of pilots lies in the optimal way to exploit all the information the receiver is provided with. Classical methods consider that only data received from pilot positions are informative. As a consequence, between pilot positions, the estimated channel must be reconstructed using interpolation techniques that would prove robust (whatever one means by robustness) in simulations \cite{COL02}. Then it appeared that a Bayesian minimum mean square error (MMSE) \cite{KAY93} can be derived when not only the pilot sequences but also the channel covariance matrix are known. This solution coincides with the linear MMSE (LMMSE) estimator and thus provides optimal performance when the state of knowledge on the system is limited to those pilot data and to the channel covariance matrix \cite{MOR01}. However, when the channel covariance matrix is unknown, then again, only \textit{ad-hoc} techniques were derived to cope with the lack of knowledge. The definite choice of a prior correlation matrix for the channel is one of the classical approach (identity matrix or exponentially decaying matrix for instance) \cite{ANC08}. But all those approaches are only justified by good performance arising in selected simulations and do not provide any proof as for their overall performance.

In the following work, we tackle channel estimation for OFDM as a problem of {\it inductive reasoning} based on the available prior information and the received pilots. Especially, to recover missing information, we extensively use the \textit{maximum entropy principle}, introduced by Shannon \cite{SHA48}, extended by Jaynes \cite{JAY79} and accurately proven by Shore and Johnson \cite{SHO80} to be the desirable mathematical tool to cope with lack of information. Some of the aforementioned classical results shall be found anew and proven optimal in our information theoretic framework, while new results will be provided which show to perform better than classical approaches.
The remainder of this paper is structured as follows: In Section \ref{sec:model}, we introduce the channel pilot-aided OFDM system, then in Section \ref{sec:chest}, we carry out the Bayesian channel estimation study based on different levels of knowledge. Simulations are then proposed in Section \ref{sec:simulation} and a thorough discussion on the results, limitations and extendability of our scheme is handled in Section \ref{sec:discussion}. Then we draw our conclusions in Section \ref{sec:conclusion}.

{\it Notations}: In the following, boldface lower case symbols represent
vectors, capital boldface characters denote matrices (${\bf I}_N$ is the
$N\times N$ identity matrix). The transposition operation is denoted $(\cdot)^{\sf T}$. The Hermitian transpose is denoted $(\cdot)^{\sf H}$. The operator $\diag({\bf x})$ turns the vector $\bf x$ into a diagonal matrix. The symbol $\det({\bf X})$ is the determinant of matrix ${\bf X}$. The symbol $\EE[\cdot]$ denotes expectation. The Kronecker delta function is denoted $\delta_{x}$ that equals $1$ if $x=0$ and equals $0$ otherwise.

\section{System Model}
\label{sec:model}
\begin{figure}
\centering
\includegraphics[width=10cm]{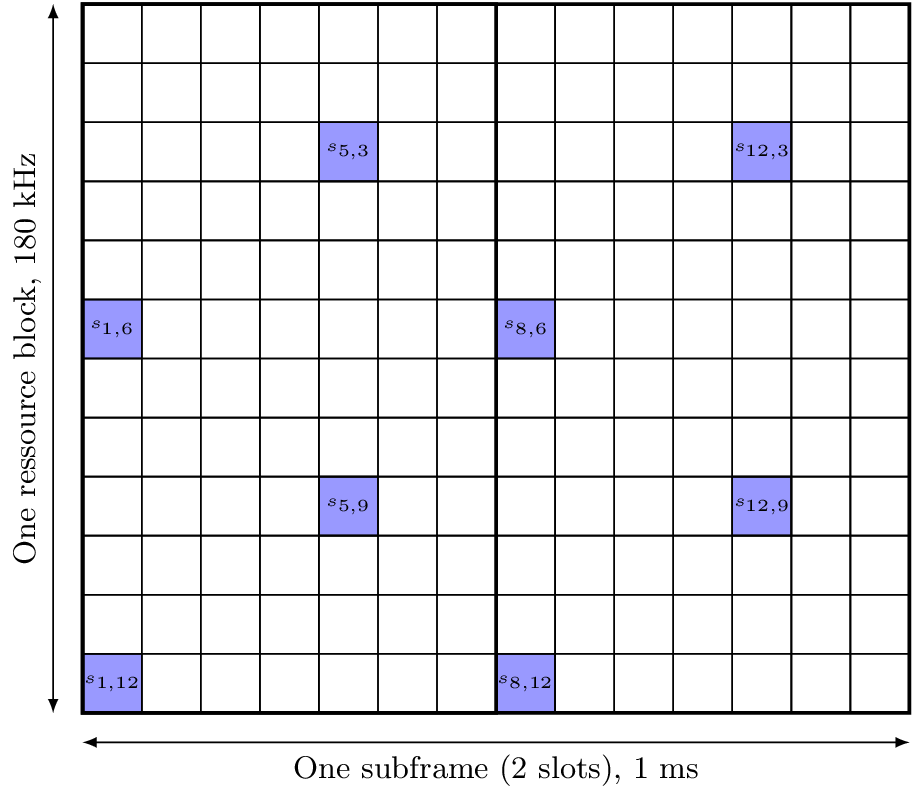}
\caption{Time-frequency OFDM grid with pilot positions enhanced}
\label{fig:pilots}
\end{figure}

We consider here a single cell OFDM system with $N$ subcarriers. The cyclic prefix (CP) length is $N_{CP}$ samples. In the time-frequency OFDM symbol grid, pilots are found in the symbol positions indexed by the function $\phi_t(n)\in \{0,1\}$ which equals $1$ if a pilot symbol is present at subcarrier $n$ and $0$ otherwise; the subscript index $t$ denotes the OFDM symbol time. This is depicted in Figure \ref{fig:pilots}. Both data and pilots are gathered at time $t$ in a frequency-domain vector ${\bf s}_t\in \mathbb C^N$ with pilot entries of amplitude $|s_{t,k}|^2=1$, for all $t,k$ such that $\phi_t(k)=1$. They are then sent through a channel of frequency response ${\bf h}_t\in \mathbb C^N$ with entries of mean power $\EE[|h_{t,k}|^2]=1$, and background noise ${\bf n}_t\in \mathbb C^N$ with entries of mean power $\EE[|n_{t,k}|^2]=\sigma^2$. The time-domain representation of ${\bf h}_t$ is denoted ${\boldsymbol \nu}_t\in \mathbb C^L$ with $L$ the channel length, i.e. the number of time-domain samples which suffer inter symbol interference. The OFDM frequency signal ${\bf y}_t\in \mathbb C^N$ received at time $t$ at the receiver is
\begin{equation}
\label{eq:model}
{\bf y}_t = \diag({\bf h}_t){\bf s}_t+{\bf n}_t
\end{equation}

This work aims at optimally estimating the vector ${\bf h}_t$ for some performance measure defined hereafter. The estimate shall be denoted $\hat{\bf h}_t\in \mathbb C^N$.

Different states of knowledge at the receiver are considered in the subsequent work,
\begin{itemize}
\item the channel length $L$ is either known or unknown,
\item at discrete time $t$, the pilots at time $(t-k)$ for $k\in \{1,\ldots,K\}$ as well as the channel time-correlation, are either known or unknown.
\end{itemize}

On top of those parameters, classical system parameters are supposed to be known (some were explicitly already used),
\begin{itemize}
\item signal mean power
\item noise mean power
\item channel mean power.
\end{itemize}

In each of the subsequent derivations, the exact quantity of knowledge will be clearly stated, since it is essential to the inductive inference we will perform on the channel ${\bf h}$.

\section{Channel Estimation}
\label{sec:chest}
\subsection{The channel length is known}
\label{sec:Lknown}
We first consider the simple scenario in which the channel power delay profile, i.e. the diagonal elements of the time-domain channel covariance matrix, is unknown but the channel length $L$ is known. Only pilot sequences received at discrete time $t$ are known to the receiver. This is, we assume that the receiver is not able to register either past received signals, nor past estimates of the channel. This hypothesis will be relieved in subsequent considerations. The amount of prior information, i.e. noise power, signal power and $L$, is denoted $I$. The amount of information that can be inferred about some entity $E$ from the prior information $I$ will be denoted $(E|I)$. For ease of reading, we will remove the index $t$ in the notations when unnecessary. We will also, $\forall k\in \{1,\ldots,N\}$, denote $h_k'=y_k/s_k=h_k+n_k/s_k$ and ${\bf h}'=(h_1',\ldots,h_N')^{\sf T}$.

The only knowledge on the additional noise vector in the system is the mean power $\sigma^2$ of its entries. The maximum entropy principle \cite{JAY03} requires then that the noise process be assigned a Gaussian independent and identically distributed (i.i.d.) density: ${\bf n}\sim C\mathcal N(0,{\sigma^2\bf I}_N)$. From the channel model \eqref{eq:model}, equivalently, the multipath channel of length $L$ is only known to be of unit mean power. Again, the maximum entropy principle demands ${\boldsymbol \nu} \sim C\mathcal N(0,\frac{1}{L}{\bf I}_L)$. Since $\boldsymbol \nu$ is a Gaussian vector with i.i.d. entries, $\bf h$, its discrete Fourier transform, is a correlated Gaussian vector,
\begin{equation} 
{\bf h} \sim C\mathcal N(0,{\bf Q})
\end{equation}
with, for any couple $(n,m)\in \{1,\ldots,N\}^2$,
\begin{align}
Q_{nm} &=\EE\left[ \sum_{k=0}^{L-1}\sum_{l=0}^{L-1}\nu_k\nu_l^{\ast}e^{-2\pi i\frac{kn-lm}{N}} \right] \\
&= \frac{1}{L}\sum_{k=0}^{L-1}e^{-2\pi ik\frac{n-m}{N}}
\end{align}

To derive the optimal channel estimator, a decision regarding the targeted error function to be minimized has to be made. Following most of previous contributions in this respect, we propose to take $\hat{\bf h}$ as the estimator that minimizes the mean quadratic estimation error (MMSE estimator), given the received signal ${\bf y}$. This is \cite{KAY93}

\begin{align}
\hat{\bf h} &= \EE\left[ {\bf h}|{\bf y}\right] \\
&= \int_{\mathbb C^N}{\bf h} \frac{P({\bf h})P({\bf y}|{\bf h})}{P({\bf y})}d{\bf h} \\
&= \int_{\mathbb C^N}{\bf h} \frac{P({\bf h})P({\bf y}|{\bf h})}{\left(\int_{\mathbb C^N}P({\bf h})P({\bf y}|{\bf h})d{\bf h}\right)}d{\bf h} \\ 
\label{eq:lim}&= \lim_{\tilde{\bf Q}\rightarrow {\bf Q}}\frac{1}{{\pi}^{N+M}\sigma^{2M}\det{\tilde{\bf Q}}}\int_{\mathbb C^N}\frac{{\bf h}\cdot e^{-{\bf h}^{\sf H}{\tilde{\bf Q}}^{-1}{\bf h}}\cdot e^{-\frac{1}{\sigma^2}({\bf h}-{\bf h}')^{\sf H}{\bf P}^{\sf H}{\bf P}({\bf h}-{\bf h}')} d{\bf h}}{\left( \frac{1}{{\pi}^{N+M}\sigma^{2M}\det{\tilde{\bf Q}}} \int_{\mathbb C^N} e^{-{\bf h}^{\sf H}{\tilde{\bf Q}}^{-1}{\bf h}}\cdot e^{-\frac{1}{\sigma^2}({\bf h}-{\bf h}')^{\sf H}{\bf P}^{\sf H}{\bf P}({\bf h}-{\bf h}')}d{\bf h}\right)} \\
\label{eq:bigeq} &= \lim_{\tilde{\bf Q}\rightarrow {\bf Q}}\int_{\mathbb C^N}\frac{{\bf h}\cdot e^{-{\bf h}^{\sf H}{\tilde{\bf Q}}^{-1}{\bf h}}\cdot e^{-\frac{1}{\sigma^2}({\bf h}-{\bf h}')^{\sf H}{\bf P}^{\sf H}{\bf P}({\bf h}-{\bf h}')} d{\bf h}}{\left( \int_{\mathbb C^N} e^{-{\bf h}^{\sf H}{\tilde{\bf Q}}^{-1}{\bf h}}\cdot e^{-\frac{1}{\sigma^2}({\bf h}-{\bf h}')^{\sf H}{\bf P}^{\sf H}{\bf P}({\bf h}-{\bf h}')}d{\bf h}\right)}
\end{align}
in which the limiting process is taken over a set of invertible matrices $\tilde{\bf Q}$ (which tends to $\bf Q$ that is by definition of rank $L<N$), and $\bf P$ is a projection matrix over the set of pilot frequency carriers ($P_{ij}=\delta_{i-j}\delta_{\phi(i)}$).

Let us note at this point that, from the received data ${\bf y}$, we only use the symbols indexed at pilot positions in equation \eqref{eq:lim}, hence the introduction of the projectors ${\bf P}$. This seems to go against our optimality claim. Indeed, one might object that data outside the pilot positions somehow carry information about the channel and should be taken into account. In the same manner, we could also claim that potential interferers are not correctly dealt with. Still, our problem is correctly formulated since $I$ does not carry any information about informative data apart from pilots, nor does it even suggest the presence of interferers. This epistemological discussion is further debated in Section \ref{sec:discussion}.

The product of the exponential terms in \eqref{eq:bigeq} can be written (by expansion and identification)
\begin{align}
\label{eq:transform}
-{\bf h}^{\sf H}{\tilde{\bf Q}}^{-1}{\bf h}-\frac{1}{\sigma^2}({\bf h}-{\bf h}')^{\sf H}{\bf P}^{\sf H}{\bf P}({\bf h}-{\bf h}') = -({\bf h}-\tilde{\bf k})^{\sf H}\tilde{\bf M}({\bf h}-\tilde{\bf k}) - \tilde{C}
\end{align}
with
\begin{equation}
\left\{
\begin{array}{ll}
\tilde{\bf M}&={\tilde{\bf Q}}^{-1}+\frac{1}{\sigma^2}{\bf P}^{\sf H}{\bf P} \\
\tilde{\bf k}&=\frac{1}{\sigma^2}\tilde{\bf M}^{-1}{\bf P}^{\sf H}{\bf P}{\bf h}' \\
\tilde{C}&=\frac{1}{\sigma^2}{{\bf h}'}^{\sf H}{\bf P}^{\sf H}{\bf P}{\bf h}'-\tilde{\bf k}^{\sf H}\tilde{\bf M}\tilde{\bf k}
\end{array}
\right.
\end{equation}

This allows to isolate the dumb variable $\bf h$ in the integrals and leads then to compute the first order moment of a multivariate Gaussian distribution,
\begin{align}
\hat{\bf h} &=\lim_{\tilde{\bf Q}\rightarrow {\bf Q}}\int_{\mathbb C^N}\frac{{\bf h}\cdot e^{-({\bf h}-\tilde{\bf k})^{\sf H}\tilde{\bf M}({\bf h}-\tilde{\bf k})}d{\bf h}}{\int_{\mathbb C^N}e^{-({\bf h}-\tilde{\bf k})^{\sf H}\tilde{\bf M}({\bf h}-\tilde{\bf k})}d{\bf h}} \\
&= \lim_{\tilde{\bf Q}\rightarrow {\bf Q}} \tilde{\bf k}
\end{align}
in which $\tilde{\bf k}$ depends on $\tilde{\bf Q}$ through $\tilde{\bf M}$.

Noting that $\tilde{\bf M}^{-1}=({\tilde{\bf Q}}^{-1}+\frac{1}{\sigma^2}{\bf P}^{\sf H}{\bf P})^{-1}=({\bf I}_N+\frac{1}{\sigma^2}{\tilde{\bf Q}}{\bf P}^{\sf H}{\bf P})^{-1}\tilde{\bf Q}$, $\hat{\bf h}$ is then
\begin{align}
\hat{\bf h} &=\lim_{\tilde{\bf Q}\rightarrow {\bf Q}}(\sigma^2{\bf I}_N+{\tilde{\bf Q}}{\bf P}^{\sf H}{\bf P})^{-1}\tilde{\bf Q}{\bf P}^{\sf H}{\bf P}{\bf h}'\\
\label{eq:estL} &= (\sigma^2{\bf I}_N+{{\bf Q}}{\bf P}^{\sf H}{\bf P})^{-1}{\bf Q}{\bf P}^{\sf H}{\bf P}{\bf h}'
\end{align}

This is exactly the well-known LMMSE solution \cite{KAY93}. However, this result is not yet another demonstration of LMMSE as it is classically derived. We used here the maximum entropy principle to observe that, at the end, the limited knowledge on the channel length $L$ mathematically gives the same estimation as when one ``assumes'' an {\it a priori} covariance matrix ${\bf Q}$. Therefore the intuitive classical solution is the correct estimate in the sense of maximum entropy \cite{JAY03}.

\subsection{Unknown channel length}
If $L$ is only known to be in an interval $\{L_{\sf min},\ldots,L_{\sf max}\}$, the maximum entropy principle assigns a uniform prior distribution for $L$; otherwise one would add non desirable implicit information to the current state of knowledge. The channel MMSE estimator is then given by
\begin{align}
\hat{\bf h} &= \EE\left[ {\bf h}|{\bf y}\right] \\
&= \int_{\mathbb C^N}{\bf h} \frac{\left(\sum_LP({\bf h}|L)P(L)\right)P({\bf y}|{\bf h})}{\left(\int_{\mathbb C^N}\left(\sum_LP({\bf h}|L)P(L)\right)P({\bf y}|{\bf h})d{\bf h}\right)}d{\bf h} \\
\label{eq:estfulleq}&= \lim_{\substack{\tilde{\bf Q}_{L_{min}}\rightarrow {\bf Q}_{L_{min}} \\ \cdots \\ \tilde{\bf Q}_{L_{max}}\rightarrow {\bf Q}_{L_{max}}}}\sum_{L=L_{min}}^{L_{max}}\frac{1}{\det{\tilde{\bf Q}_L}}\int_{\mathbb C^N}\frac{{\bf h}\cdot e^{-{\bf h}^{\sf H}{\tilde{\bf Q}}_L^{-1}{\bf h}}\cdot e^{-\frac{1}{\sigma^2}({\bf h}-{\bf h}')^{\sf H}{\bf P}^{\sf H}{\bf P}({\bf h}-{\bf h}')} d{\bf h}}{\left(\sum_{L=L_{min}}^{L_{max}}\frac{1}{\det{\tilde{\bf Q}_L}}\int_{\mathbb C^N} \cdot e^{-{\bf h}^{\sf H}{\tilde{\bf Q}}_L^{-1}{\bf h}}\cdot e^{-\frac{1}{\sigma^2}({\bf h}-{\bf h}')^{\sf H}{\bf P}^{\sf H}{\bf P}({\bf h}-{\bf h}')}d{\bf h}\right)}
\end{align}
where ${\bf Q}_k$ is the channel covariance matrix for a channel length $k\in \{L_{min},\ldots,L_{max}\}$ and $\tilde{\bf Q}_k$ are taken in a set of invertible matrices in the neighborhood of ${\bf Q}_k$.

Using the same transformations as in \eqref{eq:transform} and the fact that the numerator and denominator constants in \eqref{eq:estfulleq} do not simplify any longer, we end up with
\begin{align}
\hat{\bf h} &= \lim_{\substack{\tilde{\bf Q}_{L_{min}}\rightarrow {\bf Q}_{L_{min}} \\ \cdots \\ \tilde{\bf Q}_{L_{max}}\rightarrow {\bf Q}_{L_{max}}}}\frac{\sum_{L=L_{min}}^{L_{max}}\det(\tilde{\bf M}^{(L)}\tilde{\bf Q}_L)^{-1}e^{-\tilde{C}^{(L)}}\tilde{\bf k}^{(L)}}{\sum_{L=L_{min}}^{L_{max}}{\det(\tilde{\bf M}^{(L)}\tilde{\bf Q}_L)^{-1}}e^{-\tilde{C}^{(L)}}}
\end{align}
in which we updated our previous notations to incorporate the dependence on $L$,
\begin{equation}
\left\{
\begin{array}{ll}
\tilde{\bf M}^{(L)} &=  {\tilde{\bf Q}}_L^{-1}+\frac{1}{\sigma^2}{\bf P}^{\sf H}{\bf P} \\
&= {\tilde{\bf Q}}_L^{-1}({\bf I}_N+\frac{1}{\sigma^2}{\tilde{\bf Q}_L}{\bf P}^{\sf H}{\bf P}) \\
\tilde{\bf k}^{(L)} &=  \frac{1}{\sigma^2}({\bf I}_N+\frac{1}{\sigma^2}{\tilde{\bf Q}_L}{\bf P}^{\sf H}{\bf P})^{-1}\tilde{\bf Q}_L{\bf P}^{\sf H}{\bf P}{\bf h}' \\
\tilde{C}^{(L)} &= \frac{1}{\sigma^2}{{\bf h}'}^{\sf H}{\bf P}^{\sf H}{\bf P}{\bf h}'-\tilde{\bf k}^{\sf H}({\tilde{\bf Q}}^{-1}+\frac{1}{\sigma^2}{\bf P}^{\sf H}{\bf P})\tilde{\bf k}\\
&= \frac{1}{\sigma^2}{{\bf h}'}^{\sf H}{\bf P}^{\sf H}{\bf P}{\bf h}'-\frac{1}{\sigma^4}\left[ ({\bf I}_N+\frac{1}{\sigma^2}{\tilde{\bf Q}_L}{\bf P}^{\sf H}{\bf P})^{-1}\tilde{\bf Q}_L{\bf P}^{\sf H}{\bf P}{\bf h}' \right]^{\sf H}{\bf P}^{\sf H}{\bf P}{\bf h}' \\
&= {{\bf h}'}^{\sf H}\left( ({\bf I}_N+\frac{1}{\sigma^2}{\tilde{\bf Q}}_L{\bf P}^{\sf H}{\bf P})^{-1}\right)^{\sf H}\frac{{\bf P}^{\sf H}{\bf P}}{\sigma^2}{\bf h}'
\end{array}
\right.
\end{equation}

The determinant $\det(\tilde{\bf M}^{(L)})$ can be further developed as
\begin{align}
\det(\tilde{\bf M}^{(L)}) &= \det({\bf I}_N+\frac{1}{\sigma^2}\tilde{\bf Q}_L{\bf P}^{\sf H}{\bf P})\cdot \det({\tilde{\bf Q}_L})^{-1}
\end{align}
which entails
\begin{align}
\det(\tilde{\bf M}^{(L)}\tilde{\bf Q}_L) &= \det({\bf I}_N+\frac{1}{\sigma^2}\tilde{\bf Q}_L{\bf P}^{\sf H}{\bf P})
\end{align}

No inversion of $\tilde{\bf Q}_k$ matrices is then necessary so that the limiting process is now straightforward
\begin{align}
\label{eq:estnoL}
\hat{\bf h} &= \frac{\displaystyle\sum_{L=L_{min}}^{L_{max}}\det\left(({\bf I}_N+\frac{1}{\sigma^2}{\bf Q}_L{\bf P}^{\sf H}{\bf P})^{-1}\right)e^{-{C}^{(L)}}{\bf k}^{(L)}}{\displaystyle\sum_{L=L_{min}}^{L_{max}}\det\left(({\bf I}_N+\frac{1}{\sigma^2}{\bf Q}_L{\bf P}^{\sf H}{\bf P})^{-1}\right)e^{-{C}^{(L)}}}
\end{align}
in which ${\bf k}^{(L)}$ and $C^{(L)}$ are the limits of $\tilde{\bf k}^{(L)}$ and $\tilde{C}^{(L)}$ respectively,

\begin{equation}
\left\{
\begin{array}{ll}
{\bf k}^{(L)} &= \left({\bf I}_N+\frac{1}{\sigma^2}{{\bf Q}_L}{\bf P}^{\sf H}{\bf P}\right)^{-1}\frac{1}{\sigma^2}{\bf Q}_L{\bf P}^{\sf H}{\bf P}{\bf h}' \\ 
{C}^{(L)} &= {{\bf h}'}^{\sf H}\left( ({\bf I}_N+\frac{1}{\sigma^2}{{\bf Q}}_L{\bf P}^{\sf H}{\bf P})^{-1}\right)^{\sf H}\frac{{\bf P}^{\sf H}{\bf P}}{\sigma^2}{\bf h}'
\end{array}
\right.
\end{equation}

Since $C^{(L)}$ comprises the quadratic term ${{\bf h}'}^{\sf H}{\bf P}^{\sf H}{\bf P}{\bf h}'$, the MMSE estimation of $\bf h$ is not linear in ${\bf h}'$, therefore the LMMSE estimate in the scenario when $L$ is unknown does not coincide with the MMSE estimate. We also note that formula \eqref{eq:estnoL} is no more than a weighted function of the individual LMMSE estimates for different hypothetical values of $L$. The weighting coefficients allow to enhance the estimates that rather fit the correct $L$ hypothesis and to discard those estimates that do not concord with the ${\bf h}'$ observation.

\subsection{Using time correlation}
When the channel \textit{coherence time}, defined as the typical duration for which the channel realizations are correlated \cite{PRO98}, is of the same order or larger than a few OFDM symbols, then past (and future) received data carry important information on the present channel. This information must be taken into account.

Classically, channel time correlation is described through Jakes' model \cite{DEN93}. For a Doppler spread $f_d$ (proportional to the vehicular speed), the correlation figure is modeled as
\begin{equation}
\label{eq:jakes}
\EE[\nu_{t+T,p}\nu_{t,p}^{\ast}] = \frac{1}{L}\cdot J_0(2\pi f_dT)
\end{equation}
in which $p$ is one of the paths of the multipath channel ${\boldsymbol \nu}$ and $J_0$ is the Bessel function of the first kind. 

This model actually makes two assumptions that, under the proper information setting, can be turned into the output of a maximum entropy process. Those assumptions \cite{JAK74} are
\begin{itemize}
\item the signal scatterers are uncorrelated in the sense that two rays arriving at different angles to the receiver face independent attenuation properties. Under no knowledge on the environmental scatterers, this has to be the logical assumption.
\item the angles of arrival (i.e. the angle between the antenna body and the incoming wave) are uniformly distributed. Again, this is what the maximum entropy principle would state if no particular knowledge on the positions of the scatterers, transmitter and receiver is a priori given.
\end{itemize}

For those reasons, Jakes' model is \textit{reasonable} when no geometrical information on the channel is given. Practically speaking, it will be difficult for the receiver to be aware of the exact Doppler frequency $f_d$. In the following theoretical derivations and in the forthcoming simulations, we shall consider that the receiver exactly knows the expected value of equation \eqref{eq:jakes}. It is of course possible, either to find estimates for $\EE[\nu_{p,t+T}\nu_{p,t}^{\ast}]$ or to complete the subsequent study by integrating out the possible Doppler frequencies given a prior distribution for $f_d$.

Let us consider the simple scenario in which only the present and last past pilot symbols are considered by the terminal. Those correspond to two time instants $t_1$ and $t_2$, respectively. We also consider first that $L$ is known. For notational simplicity, we shall denote ${\bf h}_k={\bf h}_{t_k}$. The MMSE estimator for ${\bf h}_2$ under this state of knowledge is then
\begin{align}
\hat{\bf h}_2 &= \EE[{\bf h}_2|{\bf h}_1'{\bf h}_2'] \\
&= \int_{{\bf h}_2}{\bf h}_2\frac{P({\bf h}_1'{\bf h}_2'|{\bf h}_2)P({\bf h}_2)}{P({\bf h}_1'{\bf h}_2')}d{\bf h}_2 \\
&= \int_{{\bf h}_2}{\bf h}_2\frac{P({\bf h}_2)P({\bf h}_2'|{\bf h}_1'{\bf h}_2)P({\bf h}_1'|{\bf h}_2)}{P({\bf h}_1'{\bf h}_2')}d{\bf h}_2 \\
\label{eq:h1ind} &= \int_{{\bf h}_2}{\bf h}_2\frac{P({\bf h}_2)P({\bf h}_2'|{\bf h}_2)P({\bf h}_1'|{\bf h}_2)}{P({\bf h}_1'{\bf h}_2')}d{\bf h}_2 \\
&= \int_{{\bf h}_2}{\bf h}_2\frac{P({\bf h}_2)P({\bf h}_2'|{\bf h}_2)\int_{{\bf h}_1}P({\bf h}_1'|{\bf h}_2{\bf h}_1)P({\bf h}_1|{\bf h}_2)d{\bf h}_1}{P({\bf h}_1'{\bf h}_2')}d{\bf h}_2 \\
\label{eq:h2ind} &= \int_{{\bf h}_2}{\bf h}_2\frac{P({\bf h}_2)P({\bf h}_2'|{\bf h}_2)\int_{{\bf h}_1}P({\bf h}_1'|{\bf h}_1)P({\bf h}_1|{\bf h}_2)d{\bf h}_1}{P({\bf h}_1'{\bf h}_2')}d{\bf h}_2
\end{align}
in which equations \eqref{eq:h1ind} and \eqref{eq:h2ind} are verified since ${\bf h}_1$ and ${\bf h}_2$ do not bring any additional information to $({\bf h}_2'|{\bf h}_2)$ and $({\bf h}_1'|{\bf h}_1)$ respectively.

At this point, we recognize that, apart from the new term $P({\bf h}_1|{\bf h}_2)$, all the probabilities to be derived here have already been produced in the previous sections. Now, our knowledge on $({\bf h}_1|{\bf h}_2)$ is limited to equation \eqref{eq:jakes}. Burg's theorem \cite{COV91} states then that the maximum entropy distribution for $({\boldsymbol \nu}_1|{\boldsymbol \nu}_2)$ is an $L$-multivariate Gaussian distribution of mean $\lambda {\boldsymbol \nu}_2$ and variance $\frac{1}{L}(1-\lambda^2){\bf I}_L$ with $\lambda=J_0(2\pi f_dT)$. Therefore, thanks to the same linearity argument as above, the distribution of $({\bf h}_1|{\bf h}_2)$ is given by
\begin{equation}
P({\bf h}_1|{\bf h}_2) = \lim_{\tilde{\bf \Phi}\rightarrow {\bf \Phi}}\frac{1}{\pi^N\det(\tilde{\bf \Phi})}e^{-({\bf h}_1-\lambda{\bf h}_2)^{\sf H}\tilde{\bf \Phi}^{-1}({\bf h}_1-\lambda{\bf h}_2)}
\end{equation}
with
\begin{equation}
\label{eq:phiq}
{\bf \Phi}(T)=(1-\lambda^2){\bf Q}
\end{equation}

Consider first the inner integral in equation \eqref{eq:h2ind}. Similarly to above, we can express
\begin{align}
P({\bf h}_1|{\bf h}_2)P({\bf h}_1'|{\bf h}_1) &= \lim_{\tilde{\bf \Phi}\rightarrow {\bf \Phi}}
\frac{1}{\pi^{M_1+N}\sigma^{2M_1}\det(\tilde{\bf \Phi})}e^{-({\bf h}_1-\tilde{\bf k}_1)^{\sf H}\tilde{\bf M}_1({\bf h}_1-\tilde{\bf k}_1)-\tilde{C}_1}
\end{align}
with
\begin{equation}
\left\{
\begin{array}{ll}
\tilde{\bf M}_1 &= \tilde{\bf \Phi}^{-1}+\frac{1}{\sigma^2}{\bf P}_1^{\sf H}{\bf P}_1 \\
\tilde{\bf k}_1 &= \tilde{\bf M}_1^{-1}(\lambda\tilde{\bf \Phi}^{-1}{\bf h}_2+\frac{1}{\sigma^2}{\bf P}_1^{\sf H}{\bf P}_1{\bf h}_1') \\
&= ({\bf I}_N+\frac{1}{\sigma^2}\tilde{\bf \Phi}{\bf P}_1^{\sf H}{\bf P}_1)^{-1}(\lambda{\bf h}_2+\frac{1}{\sigma^2}\tilde{\bf \Phi}{\bf P}_1^{\sf H}{\bf P}_1) \\
\tilde{C}_1 &= \lambda^2{\bf h}_2^{\sf H}\tilde{\bf \Phi}^{-1}{\bf h}_2+\frac{1}{\sigma^2}{{\bf h}_1'}^{\sf H}{\bf P}_1^{\sf H}{\bf P}_1{\bf h}_1'-\tilde{\bf k}_1^{\sf H}\tilde{\bf M}_1\tilde{\bf k}_1 \\
&= \lambda^2{\bf h}_2^{\sf H}\tilde{\bf \Phi}^{-1}{\bf h}_2+\frac{1}{\sigma^2}{{\bf h}_1'}^{\sf H}{\bf P}_1^{\sf H}{\bf P}_1{\bf h}_1' - (\lambda{\bf h}_2+\frac{1}{\sigma^2}\tilde{\bf \Phi}{\bf P}_1^{\sf H}{\bf P}_1{\bf h}_1')^{\sf H}\left[\left( {\bf I}_N+\frac{1}{\sigma^2}\tilde{\bf \Phi}{\bf P}_1^{\sf H}{\bf P}_1\right)^{-1}\right]^{\sf H}\tilde{\bf \Phi}^{-1}(\lambda{\bf h}_2+\frac{1}{\sigma^2}\tilde{\bf \Phi}{\bf P}_1^{\sf H}{\bf P}_1{\bf h}_1')
\end{array}
\right.
\end{equation}
and $M_1$ is the number of pilot positions in the first pilot sequence.

Now, the integration of the part dependent on ${\bf h}_1$ gives
\begin{align}
\frac{1}{\pi^{M_1+N}\sigma^{2M_1}\det(\tilde{\bf \Phi})}\int_{{\bf h}_1}e^{-({\bf h}_1-\tilde{\bf k}_1)^{\sf H}\tilde{\bf M}_1({\bf h}_1-\tilde{\bf k}_1)}d{\bf h}_1 &= \frac{\det(\tilde{\bf M}_1)^{-1}}{\pi^{M_1}\sigma^{2M_1}\det(\tilde{\bf \Phi})}\\
&= \frac{\det({\bf I}_N+\frac{1}{\sigma^2}\tilde{\bf \Phi}{\bf P}_1^{\sf H}{\bf P}_1)}{\pi^{M_1}\sigma^{2M_1}}
\end{align}
which leads to
\begin{align}
\int_{{\bf h}_1}P({\bf h}_1|{\bf h}_2)P({\bf h}_1'|{\bf h}_1)d{\bf h}_1 &=\frac{\det({\bf I}_N+\frac{1}{\sigma^2}{\bf \Phi}{\bf P}_1^{\sf H}{\bf P}_1)^{-1}}{\pi^{M_1}\sigma^{2M_1}}e^{-C_1} 
\end{align}
with, like previously, ${\bf \Phi}$ and $C_1$ the respective limits of $\tilde{\bf \Phi}$ and $\tilde{C}_1$ in the limiting process $\tilde{\bf Q}\rightarrow {\bf Q}$.

We now need to consider the outer integral, that we shall similarly develop (not forgetting $C_1$ that depends on ${\bf h}_2$) as
\begin{align}
\frac{1}{\pi^{M_2+N}\sigma^{2M_2}\det(\tilde{\bf Q})}\int_{{\bf h}_2}e^{-({\bf h}_2-\tilde{\bf k}_2)^{\sf H}\tilde{\bf M}_2({\bf h}_2-\tilde{\bf k}_2)-\tilde{C}_2}d{\bf h}_2
\end{align}
with
\begin{equation}
\left\{
\begin{array}{ll}
\tilde{\bf M}_2 &= \tilde{\bf Q}^{-1}+\lambda^2\tilde{\bf \Phi}^{-1}-\lambda^2\left( \tilde{\bf \Phi}+\frac{1}{\sigma^2}\tilde{\bf \Phi}{\bf P}_1^{\sf H}{\bf P}_1\tilde{\bf \Phi}\right)^{-1}+\frac{1}{\sigma^2}{\bf P}_2^{\sf H}{\bf P}_2 \\
&= \tilde{\bf Q}^{-1}\left((1+\frac{\lambda^2}{1-\lambda^2}){\bf I}_N - \frac{\lambda^2}{1-\lambda^2}({\bf I}_N+\frac{1}{\sigma^2}\tilde{\bf \Phi}{\bf P}_1^{\sf H}{\bf P}_1)^{-1}+\frac{1}{\sigma^2}\tilde{\bf Q}{\bf P}_2^{\sf H}{\bf P}_2\right) \\
\tilde{\bf k}_2 &= \tilde{\bf M}_2^{-1}\left( \frac{1}{\sigma^2}{\bf P}_2^{\sf H}{\bf P}_2{\bf h}_2'+\left( \tilde{\bf \Phi}+\frac{1}{\sigma^2}\tilde{\bf \Phi}{\bf P}_1^{\sf H}{\bf P}_1\tilde{\bf \Phi}\right)^{-1}\frac{\lambda}{\sigma^2}\tilde{\bf \Phi}{\bf P}_1^{\sf H}{\bf P}_1{\bf h}_1'\right)\\
&= \left((1+\frac{\lambda^2}{1-\lambda^2}){\bf I}_N - \frac{\lambda^2}{1-\lambda^2}({\bf I}_N+\frac{1}{\sigma^2}\tilde{\bf \Phi}{\bf P}_1^{\sf H}{\bf P}_1)^{-1}+\frac{1}{\sigma^2}\tilde{\bf Q}{\bf P}_2^{\sf H}{\bf P}_2\right)^{-1}\tilde{\bf Q} \nonumber \\
&~~\times \left(\frac{1}{\sigma^2}{\bf P}_2^{\sf H}{\bf P}_2{\bf h}_2'+({\bf I}_N+\frac{1}{\sigma^2}{\bf P}_1^{\sf H}{\bf P}_1\tilde{\bf \Phi})^{-1}\frac{\lambda}{\sigma^2}{\bf P}_1^{\sf H}{\bf P}_1{\bf h}_1'\right) \\
\tilde{C}_2 &= \frac{1}{\sigma^2}{{\bf h}_2'}^{\sf H}{\bf P}_2^{\sf H}{\bf P}_2{\bf h}_2'-\tilde{\bf k}_2^{\sf H}\tilde{\bf M}_2\tilde{\bf k}_2+\frac{1}{\sigma^2}{{\bf h}_1'}^{\sf H}{\bf P}_1^{\sf H}{\bf P}_1{\bf h}_1'-{{\bf h}_1'}^{\sf H}\left[\left( {\bf I}_N+\frac{1}{\sigma^2}\tilde{\bf \Phi}{\bf P}_1^{\sf H}{\bf P}_1\right)^{-1}\right]^{\sf H}\frac{1}{\sigma^2}{\bf P}_1^{\sf H}{\bf P}_1{\bf h}_1'
\end{array}
\right.
\end{equation}
with $M_2$ the number of pilots in the second pilot sequence.

In the expression of $C_2$, it is readily seen that expanding ${\bf k}_2^{\sf H}{\bf M}_2{\bf k}_2$ leads to invert ${\bf M}_2$. The $\tilde{\bf Q}^{-1}$ factor cancels then out (from the development of ${\bf k}_2$). Then ${\bf M}_2$ and ${\bf M}_2^{-1}$ cancel out as well (in the development of ${\bf M}_2{\bf k}_2$). Therefore, no problem of matrix inversion is found in those expressions. We can then take the limit to finally have

\begin{align}
\hat{\bf h} &= \lim_{\tilde{\bf Q}\rightarrow {\bf Q}} \frac{\det(\tilde{\bf M}_2)^{-1}\det(\tilde{\bf M}_1)^{-1}}{\pi^{M_2+M_1}\sigma^{2(M_2+M_1)}\det(\tilde{\bf Q})\det(\tilde{\bf \Phi})}e^{-\tilde{C}_2}\cdot \tilde{\bf k}_2\cdot \left( \frac{\det(\tilde{\bf M}_2)^{-1}\det(\tilde{\bf M}_1)^{-1}}{\pi^{M_2+M_1}\sigma^{2(M_2+M_1)}\det(\tilde{\bf Q})\det(\tilde{\bf \Phi})}e^{-\tilde{C}_2} \right)^{-1} \\
&= \lim_{\tilde{\bf Q}\rightarrow {\bf Q}}\tilde{\bf k}_2 \\
\label{eq:solCTnoL} &= \left((1+\frac{\lambda^2}{1-\lambda^2}){\bf I}_N - \frac{\lambda^2}{1-\lambda^2}({\bf I}_N+\frac{1-\lambda^2}{\sigma^2}{\bf Q}{\bf P}_1^{\sf H}{\bf P}_1)^{-1}+\frac{1}{\sigma^2}{\bf Q}{\bf P}_2^{\sf H}{\bf P}_2\right)^{-1}{\bf Q} \nonumber \\
&~~\times \left(\frac{1}{\sigma^2}{\bf P}_2^{\sf H}{\bf P}_2{\bf h}_2'+({\bf I}_N+\frac{1-\lambda^2}{\sigma^2}{\bf P}_1^{\sf H}{\bf P}_1{\bf Q})^{-1}\frac{\lambda}{\sigma^2}{\bf P}_1^{\sf H}{\bf P}_1{\bf h}_1'\right) \\
\end{align}

This formula stands only when the channel length $L$ is known. Then, with the same notations as in previous sections, if the channel length were only known to belong to an interval $\{L_{\sf min},\ldots,L_{\sf max}\}$, then
\begin{align}
\hat{\bf h}_2 &=\left(\sum_{L=L_{\sf min}}^{L_{\sf max}}\det({\bf A}_L)^{-1}\det({\bf B}_L)^{-1} e^{{-C_2}^{(L)}}\right)^{-1} \sum_{L_{\sf min}}^{L_{\sf max}}\det({\bf A}_L)^{-1}\det({\bf B}_L)^{-1} e^{{-C_2}^{(L)}}{\bf k}_2^{(L)}
\end{align}
with
\begin{equation}
\label{eq:solCTL}
\left\{
\begin{array}{ll}
{\bf A}_L &= {\bf I}_N+\frac{1-\lambda^2}{\sigma^2}{\bf Q}_L{\bf P}_1^{\sf H}{\bf P}_1 \\
{\bf B}_L &= \left(1+\frac{\lambda^2}{1-\lambda^2}\right){\bf I}_N - \frac{\lambda^2}{1-\lambda^2}\left({\bf I}_N+\frac{1}{\sigma^2}{\bf Q}{\bf P}_1^{\sf H}{\bf P}_1\right) \\
C_2 &=  \frac{1}{\sigma^2}{{\bf h}_2'}^{\sf H}{\bf P}_2^{\sf H}{\bf P}_2{\bf h}_2'-{\bf k}_2^{\sf H}{\bf M}_2{\bf k}_2+\frac{1}{\sigma^2}{{\bf h}_1'}^{\sf H}{\bf P}_1^{\sf H}{\bf P}_1{\bf h}_1'-{{\bf h}_1'}^{\sf H}\left( {\bf I}_N+\frac{1-\lambda^2}{\sigma^2}{\bf Q}{\bf P}_1^{\sf H}{\bf P}_1\right)^{-1}\frac{1}{\sigma^2}{\bf P}_1^{\sf H}{\bf P}_1{\bf h}_1'
\end{array}
\right.
\end{equation}

The final formulas \eqref{eq:solCTnoL} and \eqref{eq:solCTL} are interesting in the sense that they do not directly carry any intuitive properties. Indeed, if we were to find an {\it ad-hoc} technique that is to ponder the relative importance of our prior information on ${\bf h}_2$, of the pilot data ${\bf h}_2'$ and of the past (or future) pilot data ${\bf h}_1'$, we would suggest a linear combination of those constraints. Our result is not linear in those constraints. However it carries the expected intuition in the limits,
\begin{itemize}
\item when $\lambda=0$, then the past and present channels are completely uncorrelated so that no information carried by the past pilots should be of any use. This is what is observed since then, equation \eqref{eq:solCTnoL} reduces to LMMSE solution \eqref{eq:estL}.
\item when $\lambda \rightarrow 1$, then
\begin{align}
\hat{h}_2 &= \left({\bf I}_N + \frac{1}{\sigma^2}{\bf Q}{\bf P}_1^{\sf H}{\bf P}_1 + \frac{1}{\sigma^2}{\bf Q}{\bf P}_2^{\sf H}{\bf P}_2 \right)^{-1}{\bf Q} \left(\frac{1}{\sigma^2}{\bf P}_2^{\sf H}{\bf P}_2{\bf h}_2'+\frac{1}{\sigma^2}{\bf P}_1^{\sf H}{\bf P}_1{\bf h}_1'\right)
\end{align}
which is again the same equation as \eqref{eq:solCTnoL} but now the past and present pilots ${\bf h}_1'$ and ${\bf h}_2'$ can be compiled into a single pilot sequence ${\bf h}_2''$ with the projector ${\bf R}_2^{\sf H}{\bf R}_2={\bf P}_1^{\sf H}{\bf P}_1+{\bf P}_2^{\sf H}{\bf P}_2$,
\begin{align}
\hat{h}_2 &= \left({\bf I}_N + \frac{1}{\sigma^2}{\bf Q}{\bf R}_2^{\sf H}{\bf R}_2\right)^{-1}{\bf Q} \left(\frac{1}{\sigma^2}{\bf R}_2^{\sf H}{\bf R}_2{\bf h}_2''\right)
\end{align}
\end{itemize}

Note also that \eqref{eq:solCTnoL} is linear in the variables ${\bf h}_1'$ and ${\bf h}_2'$, so that the final MMSE solution when $L$ is known is also the LMMSE solution.

\subsection{Time-frequency Channel Estimation}
Now, instead of merely estimating channels at times when pilot sequences are found, we can extend our scheme to estimate channels at any time position. For this, we shall in the following consider a channel ${\bf h}_{12}$ that we want to estimate, given the knowledge of pilot signals ${\bf h}_1'$ and ${\bf h}_2'$ found at positions of the respective ${\bf h}_1$ and ${\bf h}_2$ channels. We are also aware of $\lambda_1$ and $\lambda_2$, the respective time-correlation coefficients between the couples $({\bf h}_1,{\bf h}_{12})$ and  $({\bf h}_2,{\bf h}_{12})$.

We assume for brevity here that the channel length $L$ is known (this will avoid the heavy computation of some coefficients).

Using the same derivations as in the previous sections, the MMSE estimation for ${\bf h}_{12}$ is given by
\begin{align}
\hat{\bf h}_{12} &= \int_{{\bf h}_{12}}\frac{P({\bf h}_{12})\cdot \left( \int_{{\bf h}_2}P({\bf h}_2'|{\bf h}_2)P({\bf h}_2|{\bf h}_{12})d{\bf h}_2 \right)\cdot \left( \int_{{\bf h}_1}P({\bf h}_1'|{\bf h}_1)P({\bf h}_1|{\bf h}_{12})d{\bf h}_1 \right)}{P({\bf h}_1'{\bf h}_2')} d{\bf h}_{12}
\end{align}

We do not provide the complete derivation, which is identical in spirit as all the previous derivations to finally obtain,
\begin{align}
\hat{\bf h}_{12} &= \left((1+\frac{\lambda_1^2}{1-\lambda_1^2}+\frac{\lambda_2^2}{1-\lambda_2^2}){\bf I}_N -\frac{\lambda_1^2}{1-\lambda_1^2} ({\bf I}_N+\frac{1-\lambda_1^2}{\sigma^2}{\bf Q}{\bf P}_1^{\sf H}{\bf P}_1)^{-1} - \frac{\lambda_2^2}{1-\lambda_2^2} ({\bf I}_N+\frac{1-\lambda_2^2}{\sigma^2}{\bf Q}{\bf P}_2^{\sf H}{\bf P}_2)^{-1} \right)^{-1}{\bf Q} \nonumber \\
&~~\times \left(  \lambda_1 ({\bf I}_N+\frac{1-\lambda_1^2}{\sigma^2}{\bf Q}{\bf P}_1^{\sf H}{\bf P}_1)^{-1} \frac{1}{\sigma^2}{\bf P}_1^{\sf H}{\bf P}_1{\bf h}_1' + \lambda_2({\bf I}_N+\frac{1-\lambda_2^2}{\sigma^2}{\bf Q}{\bf P}_2^{\sf H}{\bf P}_2)^{-1} \frac{1}{\sigma^2}{\bf P}_2^{\sf H}{\bf P}_2{\bf h}_2' \right)
\end{align}
which generalizes equation \eqref{eq:solCTnoL}.

This is further generalized for a given number $K$ of pilot signals sequences ${\bf h}_k'$, $k\in \{1,\ldots,K\}$, (sent through channel ${\bf h}_k$) and a channel ${\bf h}$ (with inverse Fourier transform ${\boldsymbol \nu}$) which satisfies
\begin{align}
  \forall (i,k)\in \{1,\ldots,L\}\times\{1,\ldots,K\},~\EE[\nu_{i,t}\nu_{i,t+k}^{\ast}]= \frac{\lambda_k}{L}
\end{align}

The maximum entropy principle in this situation gives the MMSE estimator for ${\bf h}$,
\begin{align}
\hat{\bf h} &= \left((1+\sum_{k=1}^K\frac{\lambda_k^2}{1-\lambda_k^2}){\bf I}_N - \sum_{k=1}^K \frac{\lambda_k^2}{1-\lambda_k^2} ({\bf I}_N+\frac{1-\lambda_k^2}{\sigma^2}{\bf Q}{\bf P}_k^{\sf H}{\bf P}_k)^{-1} \right)^{-1}{\bf Q} \nonumber \\
&~~\times \left(  \sum_{k=1}^K\lambda_k ({\bf I}_N+\frac{1-\lambda_k^2}{\sigma^2}{\bf Q}{\bf P}_k^{\sf H}{\bf P}_k)^{-1} \frac{1}{\sigma^2}{\bf P}_k^{\sf H}{\bf P}_k{\bf h}_k'\right)
\end{align}

\subsection{Unknown correlation factor $\lambda$}
The reader might object at this point that the prior knowledge either of the vehicular speed or of the mean correlation factor $\lambda$ might only be accessible through yet another estimation process. Moreover $\lambda$ expresses as an expectation so that possibly some time is required to have an accurate estimation. This of course goes against our idea of fast channel estimation.

In the same trend as we did previously with the possibly unknown parameter $L$, we can equally integer out the parameter $\lambda$ from our formulas. In general this might be an uneasy task, since then the estimated channel $\hat{\bf h}$ would read,
\begin{align}
  \hat{\bf h} &= \int_{\bf h}{\bf h}P({\bf h}|{\bf h}_1',\ldots ,{\bf h}_K'){\rm d}{\bf h} \\
  &= \int_{\bf h}\left(\int_{\lambda}{\bf h}P({\bf h}|{\bf h}_1',\ldots ,{\bf h}_K',\lambda)P(\lambda)d{\lambda}\right){\rm d}{\bf h}
\end{align}
in which $P(\lambda)$ is our prior knowledge on the parameter $\lambda$. But, going further in the computation, this last integration is rather involved. It could then be well approximated by the finite summation,
\begin{align}
\hat{\bf h} &\simeq \int_{\bf h}\left(\sum_{\lambda = \lambda_{min}}^{\lambda_{max}}{\bf h}P({\bf h}|{\bf h}_1',\ldots ,{\bf h}_K',\lambda)P(\lambda)\right)d{\bf h} \\
&= \sum_{\lambda = \lambda_{min}}^{\lambda_{max}}P(\lambda) \int_{\bf h} {\bf h}P({\bf h}|{\bf h}_1',\ldots ,{\bf h}_K',\lambda)d{\bf h}
\end{align}

As shall be illustrated in Section \ref{sec:simulation}, it actually makes almost no difference to assume that $\lambda$ is or is not known. This suggests that our estimators are able themselves to cope with the lack of information concerning $\lambda$, just by inductive inference on $\lambda$ from the data ${\bf h}_1',\ldots ,{\bf h}_K'$.

\subsection{Non-homogeneous $SNR$}
In all previous sections, we considered an homogeneous noise power $\sigma^2$ over the frequency bandwidth. This situation is usually far from real (and often far from the actual knowledge the receiver has on the noise correlation matrix). Typically, in the presence of strong interferers working on selective frequencies, the noise correlation matrix ${\bf C}_{\bf n}=\EE[{\bf n}{\bf n}^{\sf H}]$ is not proportional to an identity matrix (and has even no particular tendency to be diagonal).

If the information about the noise is updated to consider ${\bf C_n}$, then all the previous equations are to be updated by replacing all terms $\frac{1}{\sigma^2}{\bf P}^{\sf H}{\bf P}$ by the corrected terms ${\bf P}^{\sf H}{\bf C_n}{\bf P}$.

\section{Simulations and Results}
\label{sec:simulation}
\begin{figure}
	\centering
	\includegraphics[width=10cm]{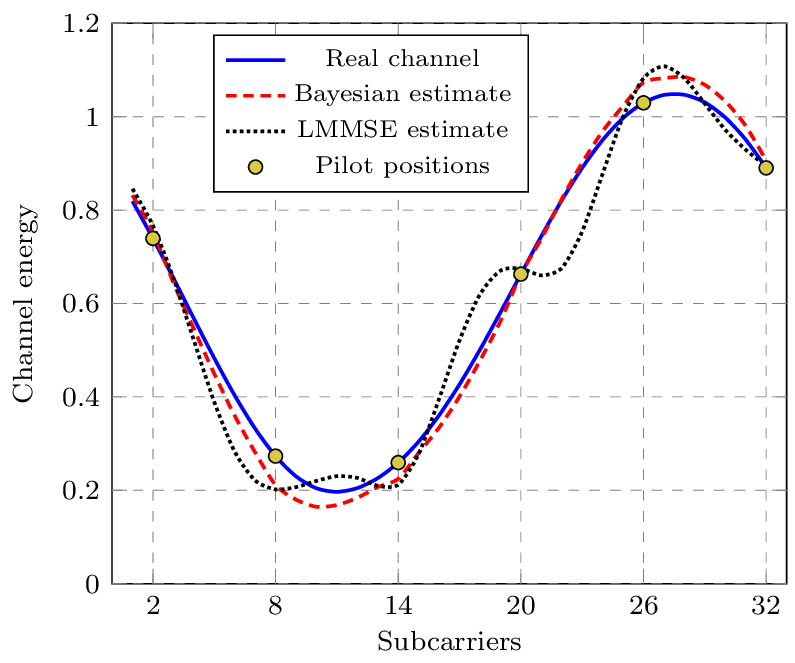}
	\caption{Channel estimated energy - $N=32$, $L=3$, $L_{\sf max}=6$, ${\sf SNR}=20~{\rm dB}$}
	\label{fig:chestL}
\end{figure}

In this section, we propose simulations of the previously derived formulas. In order to produce insightful plots, we consider a short discrete Fourier transform of size $N=32$. In a first simulation, we assume the channel length $L=3$ is unknown to the receiver that only knows $L\in \{1,\ldots,6\}$. The pilot symbols are separated by 6 subcarriers as in the 3GPP-LTE standard \cite{LTE} and depicted in Figure \ref{fig:pilots}. The SNR is set to ${\sf SNR}=20~{\rm dB}$. The results are proposed in Figure \ref{fig:chestL} that compares the novel Bayesian MMSE solution to the classical LMMSE solution assuming maximum channel length $\tilde{L}=L_{\sf max}=6$. It is observed that our solution has better results than LMMSE in this particular case, due to the prior error in $\tilde{L}$ that does not occur in this novel scheme.

\begin{figure}
	\centering
	\includegraphics[width=10cm]{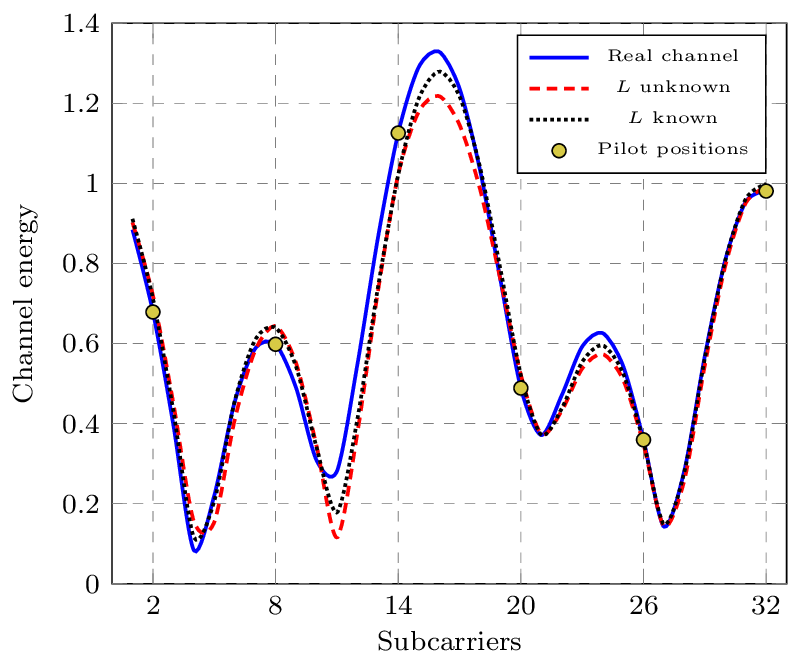}
	\caption{Channel estimated energy - $N=32$, $L=5$, $L_{\sf max}=10$, ${\sf SNR}=20~{\rm dB}$}
	\label{fig:chestL2}
\end{figure}

A comparison between the performance of our Bayesian MMSE estimator when the channel length $L$ is known or is unknown is then proposed in Figure \ref{fig:chestL2}. We take here a channel length $L=5$ that is known to belong to the range $\{1,\ldots,10\}$. Interestingly, the performance decay due to the absence of knowledge in $L$ is not large. If we consider the previous situation, i.e. $L=3$ for a range $\{1,\ldots,6\}$, we even visually observe no performance difference.

\begin{figure}
	\centering
	\includegraphics[width=10cm]{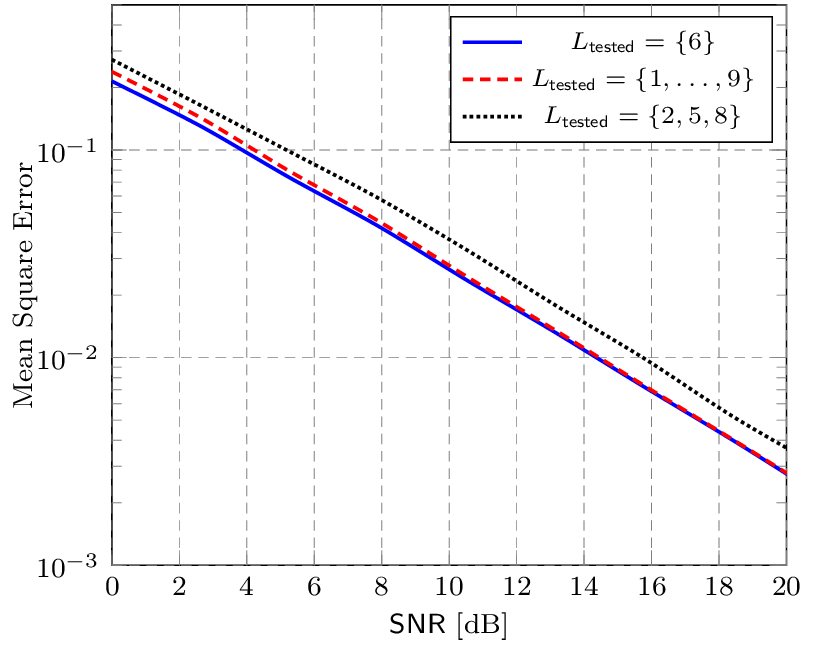}
	\caption{Mean square error of channel estimation when $L$ is unknown - ${\sf SNR}=20~{\rm dB}$, $N=128$, $L=6$}
	\label{fig:perf128}
\end{figure}

This raises a very interesting feature of the inductive reasoning framework since, by trying to infer on the channel knowledge given the received signal $({\bf h}|{\bf y})$, the Bayesian framework also encompasses inference on $L$. Indeed,
\begin{align}
P(L|{\bf y}) &=\frac{P(L)P({\bf y}|L)}{P({\bf y})} \\
&= \frac{P(L)}{P({\bf y})}\int_{\bf h}P({\bf y}|{\bf h})P({\bf h}|L){\rm d}{\bf h}
\end{align}
in which the integral is the same has in Section \ref{sec:Lknown} and $P(L)$ is the uniform prior distribution for $L$.

The inductive inference on every hypothesis $L=1,\ldots,L_{max}$ can then be compared thanks to the \textit{evidence} function defined by Jaynes \cite{JAY03} which reads
\begin{equation}
e(L|{\bf y})=\log_{10}\left( \frac{P(L|{\bf y})}{\sum_{l\neq L}P(l|{\bf y})}\right)
\end{equation}

\begin{figure}
	\centering
	\includegraphics[width=10cm]{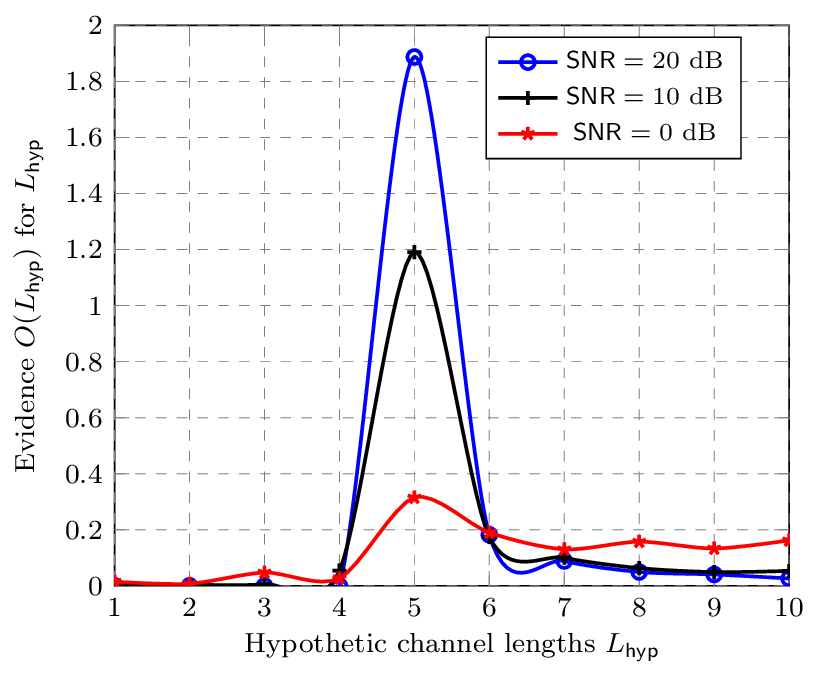}
	\caption{Channel length inference $O(L)=\frac{P(L|{\bf y})}{\sum_{l\neq L} P(l|{\bf y})}$ for different SNR - $N=32$, $L=5$, $L_{\sf max}=10$}
	\label{fig:Lupdate}
\end{figure}

The results, for different SNR are proposed in Figure \ref{fig:Lupdate}. However, since the \textit{evidence} for any hypothesis on $L$ is not large, we instead draw the \textit{odds} function
\begin{equation}
O(L|{\bf y})=\left( \frac{P(L|{\bf y})}{\sum_{l\neq L}P(l|{\bf y})} \right)
\end{equation} 
which is ten to the power of the \textit{evidence} function.

We observe, as predicted, that evidence for $L=5$ is raised higher than the other hypothesis and that this behaviour is especially noticeable when the SNR is high. This produces an updated posterior distribution $P(L|{\bf y})$ that almost discards all wrong hypothesis. Therefore, at high SNR, the impact of the hypothesis $L\neq 5$ in the final equations is negligible. This automatic inference on the channel length is a direct consequence of the proposed MMSE formula that would not be possible through classical orthodox probability approaches.

\begin{figure}
	\centering
	\includegraphics[width=10cm]{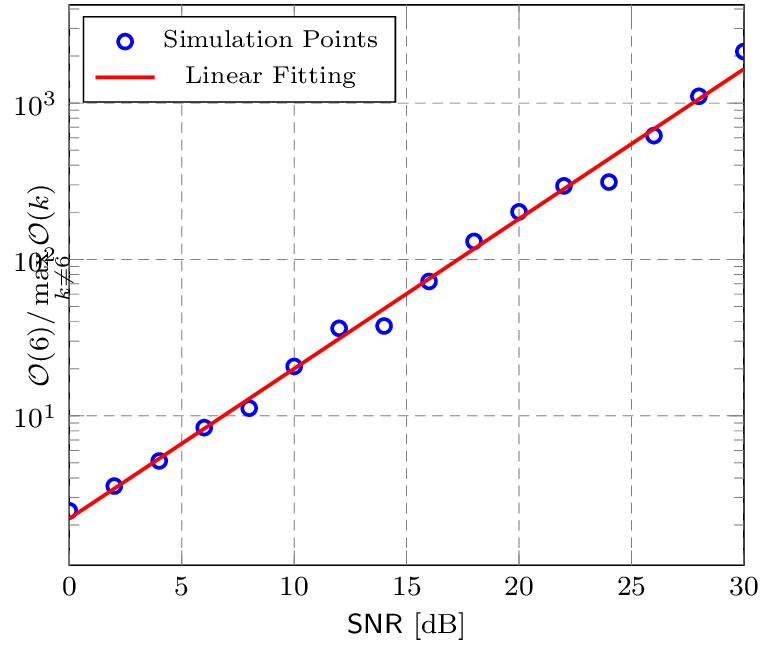}
	\caption{Mean evidence gap of channel length recovery - ${\sf SNR}=20~{\rm dB}$, $N=128$, $L=6$, $L_{\sf tested}\in \{1,\ldots,9\}$}
	\label{fig:perf_est_L_128}
\end{figure}

\begin{figure}
	\centering
	\includegraphics[width=10cm]{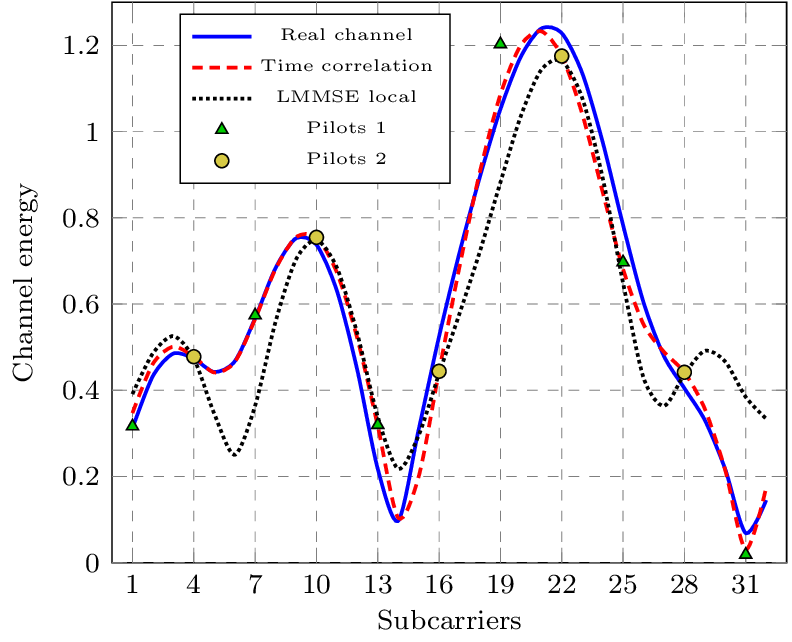}
\caption{Channel estimated energy with time correlation - $99\%$ correlation, $N=32$, $L=6$, ${\sf SNR}=20~{\rm dB}$}
\label{fig:chestTC}
\end{figure}

\begin{figure}
	\centering
	\includegraphics[width=10cm]{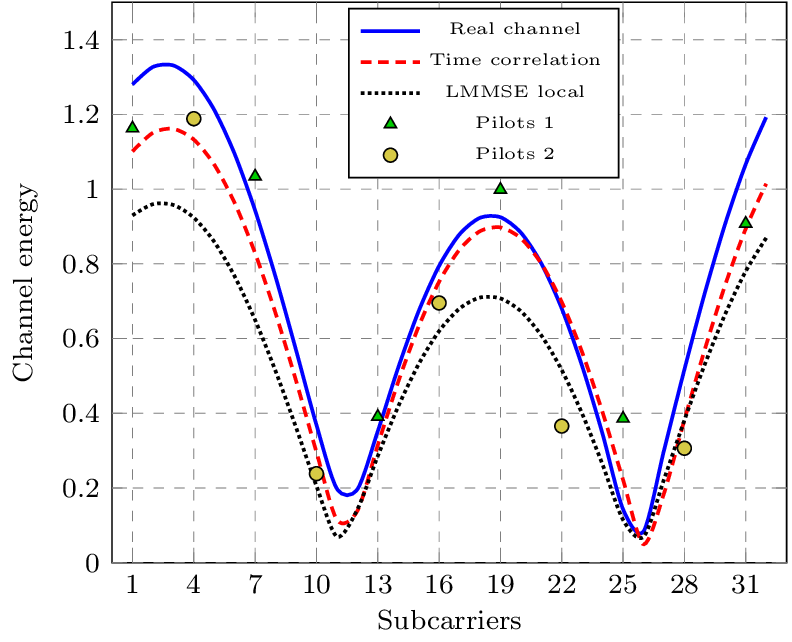}
	\caption{Channel estimated energy with time correlation - $99\%$ correlation, $N=32$, $L=3$, ${\sf SNR}=10~{\rm dB}$}
	\label{fig:chestTC2}
\end{figure}

We want in the following to observe first the effects of using time correlation properties before dealing with performance figures. In Figures \ref{fig:chestTC} and \ref{fig:chestTC2}, we propose the situation of two pilot sequences corresponding to two correlated channels with correlation factor $\lambda=0.99$. We estimate here one of the two channels either using both pilot sequences. It is observed that the high correlation $\lambda$ allows to perform the estimation of long channels. Indeed, the high density of pilots in the time-frequency grid allows to better approach the genuine channel. This is shown in Figure \ref{fig:chestTC} on a random channel, for $N=32$, $L=6$, ${\sf SNR}=20~{\rm dB}$, which clearly illustrates that using both pilot signals increases the accuracy of the estimator.

Another good effect is observed when the noise power $\sigma^2$ is strong (or equivalently when the SNR is low). Then, using twice as many samples for channel estimation leads effectively to twice as accurate channel estimation provided that both channels are strongly correlated. This is demonstrated in Figure \ref{fig:chestTC2} in which we used ${\sf SNR}=10~{\rm dB}$.

\begin{figure}
	\centering
	\includegraphics[width=10cm]{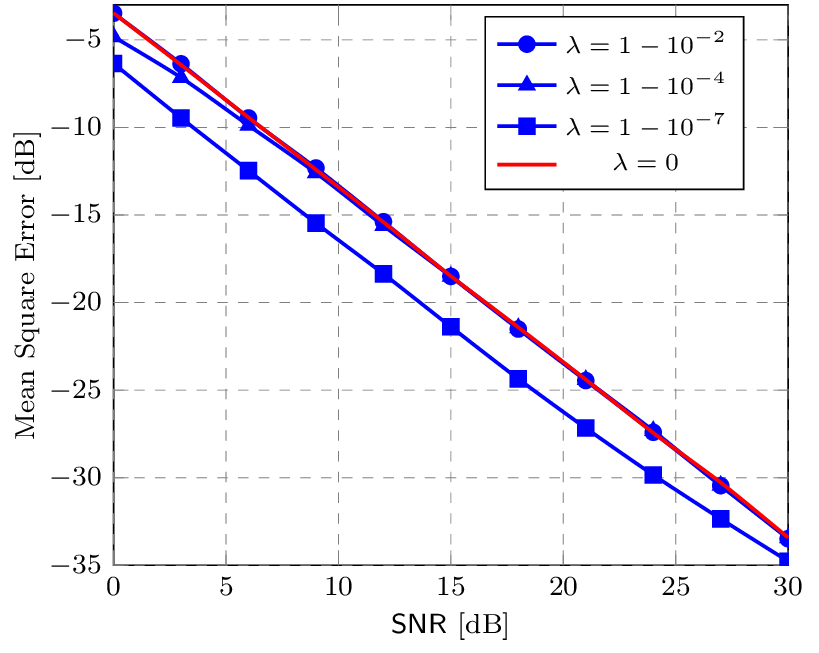}
	\caption{Estimation Mean Square Error - Short channel - $N=64$, $L=5$}
	\label{fig:chestLambdas5}
\end{figure}

\begin{figure}
	\centering
	\includegraphics[width=10cm]{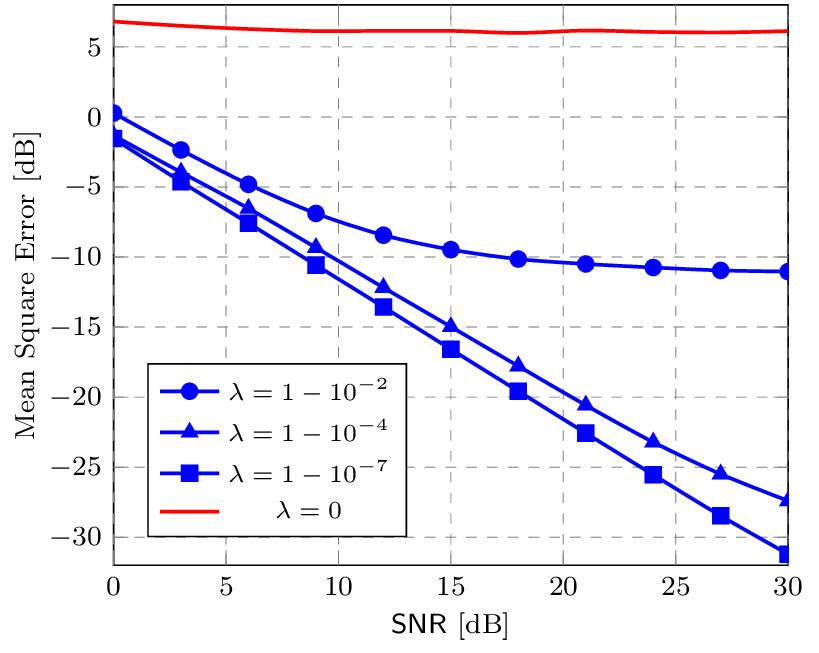}
	\caption{Estimation Mean Square Error - Long channel - $N=64$, $L=15$}
	\label{fig:chestLambdas15}
\end{figure}

The corresponding numerical performance are proposed in Figures \ref{fig:chestLambdas5} and \ref{fig:chestLambdas15} respectively. In the former, the channel length is $L=5$ for a DFT-size $N=64$. A single train of pilots is then enough to estimate the channel. However, as previously discussed, in the presence of highly time-correlated channels, the estimation noise can be reduced using both trains of pilots. Essentially, it is observed here that, since the number of pilots for both sequences is almost equal, the channel estimation based on both trains of pilots is realized on twice as many pilot positions. Therefore, when the time-correlation $\lambda$ is high, for low SNR, up to $3~{\rm dB}$ gain can be observed.

As for long channels, it is clear in Figure \ref{fig:chestLambdas15} that high correlation between channels in time is demanded so to perform accurate channel estimates. Indeed, while not using time-adjacent channels prove devastating (over $6~{\rm dB}$ of mean square error), high time-correlations allows to significantly reduce the mean square error.

\begin{figure}
	\centering
	\includegraphics[width=10cm]{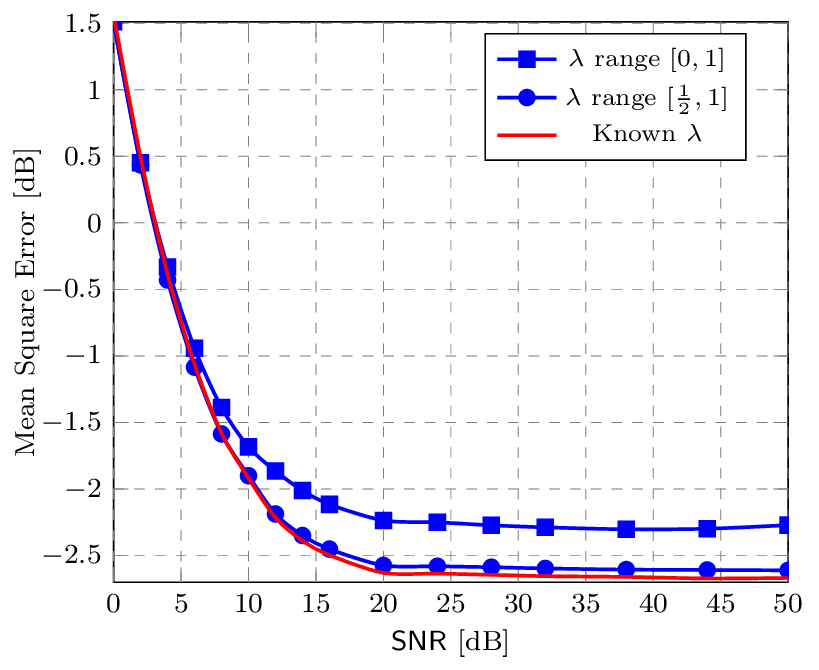}
\caption{Estimation Mean Square Error - Unknown time correlation $\lambda=93\%$ - $N=64$, $L=15$}
\label{fig:chestNLambdas}
\end{figure}

Finally, in Figure \ref{fig:chestNLambdas}, we propose to compare the performance of a channel estimator using two pilot sequences correlated in time, provided that the process knows or does not know the exact time-correlation coefficient $\lambda$, set here to $\lambda=93\%$. The channel length $L=15$ is known and the DFT size is kept to $N=64$. Surprisingly, a poor prior knowledge on $\lambda$ does not strongly impact the final mean quadratic error of the estimation. At least, when the correlation is known to be more than $1/2$, the performance in terms of mean square error is similar to that when $\lambda$ is perfectly known. This suggests, as already mentioned, that the Bayesian machinery is able to efficiently infer on $\lambda$ whatever the SNR level.
The reader must be wary that this last sentence does not imply at all that the time-correlation parameter does not intervene in the system performance. What we stated above is just that prior knowledge about this parameter is not mandatory since the posterior distribution for $\lambda$ (given ${\bf h}_1',~{\bf h}_2'$) is already very peaky around the correct value for $\lambda$.

\section{Discussion}
\label{sec:discussion}
In the following, we discuss the advantages of the general framework which encompasses the previous channel estimators. Some limitations, concerning complexity mainly, are also considered. Finally we discuss the potential drawbacks in using alternative channel estimation techniques.

Those channel estimation algorithms were proposed on the sole basis of Jaynes' probability theory \cite{JAY03} that mainly encompasses the Bayesian rule and the maximum entropy principle\footnote{the Bayesian rule has actually been proven to be a particular case of the generalized ME principle \cite{CAT08}}. Those rules can be applied to larger problems than the mere scope of channel estimation. For instance, optimal Bayesian signal detection is proposed in \cite{COU08}. Also, maximum entropy channel modeling are derived in \cite{GUI06}. All those studies lead to the general idea of \textit{cognitive receivers}. Indeed, a few years after the introduction of the concept of cognitive radios \cite{MIT99}, some attempts have been proposed to clearly define the fundamental basis of a cognitive radio \cite{HAY05} but still no correct definition has been derived. This has the rather unpleasant consequence to see many contributions on cognitive techniques, each based on very different fundamental assumptions. In our minds, probability theory as extended logic is an interesting candidate in the information theoretic definition of a cognitive receiver. By the latter, we mean a receiving terminal which, given any amount of information $I$ is able to optimally infer on any system parameter. This way the device would be capable of learning. Having done that, it will then have to take decisions, i.e. what information to send back with which accuracy, what additional information to request etc., so to maximize some utility function. However, this second requirement for a cognitive receiver goes beyond the mere scope of Jaynes' probability theory and involves decision theoretic discussions as well as epistemological considerations such as the brilliant {\it inquiry theory} from Cox \cite{COX79} and Knuth \cite{KNU02}.

This being said, the reader will then raise the objection that our current work obviously did not consider all the information available to the receiver. Indeed, as previously mentioned, if pilots were designed to help channel estimation then informative data\footnote{by informative, we suggest here data dedicated to communication purposes and not synchronization purposes} also carry information on the channel they face. We could mentally envision a joint channel estimator and signal decoder device that would help recover a better estimate on the channel and on the sent signal. But of course, at this point, we would face strong implementation difficulties as well probably as involved mathematical problems. Still, through this study we suggest that in general \textit{ad-hoc} solutions are not the direction to head for when one wishes to have both good performance and non-involved algorithms. If one wishes to reduce the computational load of an algorithm, it is preferable to develop the full Bayesian solution and \textit{only then} start to simplify the mathematical expressions. For instance, in our previous examples, when the channel length $L$ is unknown but the complexity of summing up hundreds of potentials candidates for $L$ is too large, a solution that only considers ten sampled values for $L$ could be envisioned. This is the correct way to keep a grasp on what simplification we performed; in classical techniques, from the very beginning, strong assumptions and approximations are made which effects are often invisible in terms of performance (and might only be observable through extensive simulations).

One of those classical techniques in the channel estimation realm is proposed, among others, in \cite{SHA03}. The basic idea is very insightful since it consists in estimating present channels based on the knowledge of the estimation done on a past channel and the time-correlation coefficient that link both channels. Such solutions might seem interesting in the fact that, by recursion, the previous estimate ``carries the information on all past pilots signals'' but this is actually very deluding. Indeed, this estimate does not actually contain the whole information on the previous pilot signals. It merely consists in some post-filtering result of those pilots signals. We might then raise a few objections to using previous estimates,
\begin{itemize}
  \item if we were to ``select'' information\footnote{which is dishonest according to Jaynes' theory fundamental desiderata \cite{JAY03}}, then we would better want to consider some of the past (and possibly future) pilot signals than previous estimates
\item if only the previous estimates are available, then it would seem dishonest not to mention to the Bayesian machinery that those actually \textit{are} estimates of a channel. This suggests that, when using those previous estimates, it must be somehow mentioned in the equations that they are MMSE or least-square \cite{KAY93} estimates of the channel, that originated from pilots sequences present (if this is also known) at particular positions and so on and so forth. This would allow for the Bayesian process to provide inductive inference on the actual pilot signals and then make the similar derivations as those we have proposed along this paper. Any other usage of previous estimates could not be claimed \textit{optimal}.
\end{itemize}


\section{Conclusion}
\label{sec:conclusion}
In this work, a novel view on channel estimation for OFDM systems is proposed. Optimal formulas for the mean square error criterion are derived that re-demonstrate known classical solutions while new formulas are also proposed for scenarios based on different levels of knowledge. The whole work can be summarized as a unique novel framework that allows to integrate any information the receiver is aware of to perform optimal channel inference based on this knowledge. Also, some hints on the long-term introduction of foundations for cognitive receivers are suggested that would encompass the present work.

\end{document}